\documentclass[12pt]{article}
\usepackage{amsmath,axodraw,cite}

\begin{document}

\begin{titlepage}
\begin{flushright}
UM-TH-98-03\\
February 1998\\
\end{flushright}
\vskip 2cm
\begin{center}
{\Large \bf Hadronic Contributions to the Muon Lifetime} \\[20mm]
  Timo van Ritbergen\;  and\; Robin G. Stuart \\ [2mm]
\vskip 0.5cm {\it Randall Physics Laboratory, University of Michigan\\
                  Ann Arbor, MI 48109-1120, USA}\\
\end{center}
\vskip 3cm
\hrule
\begin{abstract}
Hadronic corrections to the muon lifetime are calculated in the Fermi
theory in the presence of QED using dispersion relations.
The result, after convolution of hadron data with the calculated
perturbative kernel is
\[
\Delta\Gamma_{{\rm had}}=-\Gamma_0\left(\frac{\alpha}{\pi}\right)^2
0.042
\]
where $\Gamma_0$ is the tree-level width. The results are also used to obtain
the corrections to the muon lifetime coming from virtual muon and tau loops
\begin{eqnarray*}
\Delta\Gamma_{{\rm muon}}&=&\Gamma_0\left(\frac{\alpha}{\pi}\right)^2
\left(\frac{16987}{576}-\frac{85}{36}\zeta(2)
                        -\frac{64}{3}\zeta(3)\right)\\
                         &=&-\Gamma_0\left(\frac{\alpha}{\pi}\right)^2
                            0.0364333\\
\Delta\Gamma_{{\rm tau}}&=&-\Gamma_0\left(\frac{\alpha}{\pi}\right)^2
0.00058.
\end{eqnarray*}

\end{abstract}
\hrule
\vspace{2mm}
\noindent {
\tiny PACS: 13.35.Bv, 12.15.Lk, 12.20.Ds, 14.60.Ef }

\end{titlepage}

\setcounter{footnote}{0} \setcounter{page}{2} \setcounter{section}{0}
\newpage

\section{Introduction}

The Fermi coupling constant, $G_F$, with a current error of
$\delta G_F/G_F = 1.7\times10^{-5}$
is amongst the best measured constants in electroweak physics
and, as such, plays a crucial r\^ole as input into calculations
of electroweak observables. Of the quoted error,
$0.9\times10^{-5}$ is experimental and $1.5\times10^{-5}$ is theoretical.
The latter is an estimate of the missing and unknown higher order
QED corrections to the formula that relates the muon lifetime,
$\tau_\mu\equiv\Gamma_\mu^{-1}$, to $G_F$. The calculation of these
missing corrections will not only reduce the present error on $G_F$
by half but it also means that future experimental determinations
will be unhindered by theoretical limitations down the level
of a few parts in $10^8$.
Experiments are under consideration both at Brookhaven National Laboratory
and the Rutherford-Appleton Laboratory that could lead to a reduction
in the experimental error on the muon lifetime of about a factor
of 10. At this level, the experimental error on the muon
mass, currently $\delta m_\mu/m_\mu=3.2\times10^{-7}$, begins to become
a factor in determining the accuracy with which $G_F$ can be extracted.
It it is likely, however, that this will also undergo sufficient
improvement such that the overall error will still be dominated by that
of $\tau_\mu$.

Such an improvement in both theory and experiment would be timely since
the mass of the $Z^0$ boson $M_Z$ has been determined with an
unexpectedly high accuracy
of $\delta M_Z/M_Z=2.2\times10^{-5}$ \cite{LEPEWWG}
following the LEP runs at the $Z^0$ peak and further improvement
might still be possible. This now approaches the accuracy of $G_F$.

The 1-loop QED contributions to the muon lifetime were first calculated by
Kinoshita and Sirlin \cite{KinoSirl} and by Berman \cite{Berman}.
It is known \cite{BermSirl} that the Fermi theory in the presence
of QED is finite to first order in the Fermi
coupling constant, $G_F$, and to all orders in the electromagnetic coupling
constant, $\alpha$. This remarkable fact means that $G_F$ can be
defined in a physically unambiguous manner at least up to the point
where finite $W$ propagator effects begin to appear.

The missing corrections that affect the extraction of $G_F$ from
measurements of the muon lifetime are 2-loop QED corrections to the
Fermi theory. To these must be added single and double
bremsstrahlung contributions in order to produce infrared finite results.
Technical developments in the calculation of multiloop diagrams seem to
make the calculation of the full set of such corrections feasible.

In the present paper, corrections arising from the
hadronic vacuum polarization of the photon are considered.
These form an independent subclass of corrections with no
associated bremsstrahlung. Hadronic corrections have been calculated for
the anomalous magnetic moment of the muon
\cite{GourdeRa} and initial state corrections in
$e^+e^-\rightarrow\mu^+\mu^-$ at high energies \cite{KnieKrawKuhnStua},
by convolution of a perturbative kernel with hadronic data.
For these processes, however, the fermions
to which the virtual photons are attached have fixed 4-momenta. In the
case of muon decay, the electron participates in a phase integration
and consequently adds significantly to the complexity of the problem.

The contributions of the type that we study in the present paper were
discussed in the context of the full electroweak theory by
Sirlin \cite{Sirlin84} who demonstrated that they do not produce large
logarithms other than those that can be incorporated into the
renormalization of the electromagnetic coupling constant, $\alpha$.

\begin{figure}
\begin{center}
\begin{picture}(62,120)(0,0)
\ArrowLine(0,0)(14,40)   \Vertex(14,40){1}
\ArrowLine(14,40)(21,60)
\ArrowLine(21,60)(14,80) \Vertex(14,80){1}
\ArrowLine(14,80)(0,120)
\ArrowLine(42,120)(23,62)
\ArrowLine(23,62)(62,120)
\PhotonArc(21,60)(21.08,109.29,250.71){3}{6.5}
\GOval(-1.08,60)(5,5)(0){0.4}
\Text(8,0)[bl]{$\mu^-$}
\Text(5,122)[tl]{$e^-$}
\Text(38.5,120)[tr]{$\bar\nu_e$}
\Text(71,116)[tr]{$\nu_\mu$}
\Text(21,-3)[tl]{(a)}
\end{picture}
\qquad
\begin{picture}(62,120)(0,0)
\ArrowLine(0,0)(5.25,15)       \Vertex(5.25,15){1}
\ArrowLine(5.25,15)(15.75,45)  \Vertex(15.75,45){1}
\ArrowLine(15.75,45)(21,60)
\ArrowLine(21,60)(0,120)
\ArrowLine(42,120)(23,62)
\ArrowLine(23,62)(62,120)
\PhotonArc(10.5,30)(15.89,70.71,250.71){3}{6.5}
\GOval(-4.65,34.80)(5,5)(0){0.4}
\Text(-7,65)[bl]{1} \Text(-7,60)[bl]{--} \Text(-7,53)[bl]{2}
\Text(21,-3)[tl]{(b)}
\end{picture}
\qquad
\begin{picture}(62,120)(0,0)
\ArrowLine(0,0)(21,60)
\ArrowLine(21,60)(15.75,75)     \Vertex(15.75,75){1}
\ArrowLine(15.75,75)(5.25,105)  \Vertex(5.25,105){1}
\ArrowLine(5.25,105)(0,120)
\ArrowLine(42,120)(23,62)
\ArrowLine(23,62)(62,120)
\PhotonArc(10.5,90)(15.89,109.29,289.29){3}{6.5}
\GOval(-4.65,85.2)(5,5)(0){0.4}
\Text(-7,65)[bl]{1} \Text(-7,60)[bl]{--} \Text(-7,53)[bl]{2}
\Text(21,-3)[tl]{(c)}
\end{picture}
\qquad
\begin{picture}(62,120)(0,0)
\ArrowLine(0,0)(10.5,30)    \Vertex(10.5,30){1}
\ArrowLine(10.5,30)(21,60)
\ArrowLine(21,60)(0,120)
\ArrowLine(42,120)(23,62)
\ArrowLine(23,62)(62,120)
\Line(6.0,27.83)(15.01,32.17)
\Line(8.33,34.51)(12.67,25.50)
\Text(0,65)[bl]{1} \Text(0,60)[bl]{--} \Text(0,53)[bl]{2}
\Text(16,27)[l]{$\delta m_\mu$}
\Text(21,-3)[tl]{(d)}
\end{picture}
\\
\end{center}
\caption{Hadronic contributions to muon decay after Fierz
           rearrangement of the contact interaction.}
\end{figure}
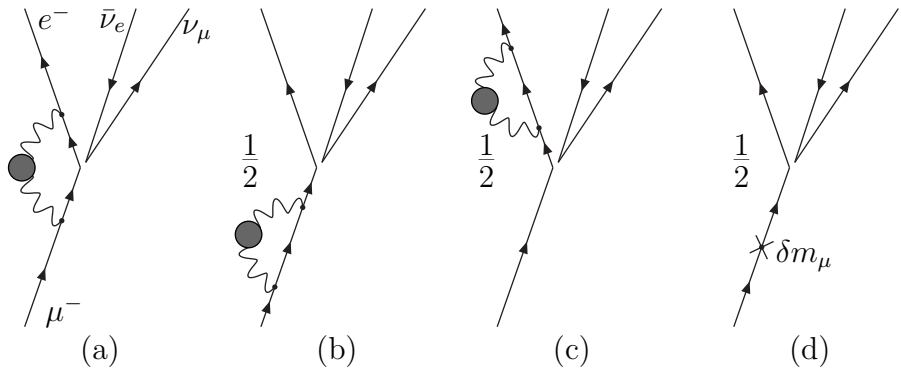

\section{Notation and Conventions}

The calculation is performed using the Euclidean metric with time-like
momenta squared being negative. The 4-momentum of the initial-state muon
will be denoted $p_\mu$ and that of the outgoing electron by $p_\mu^\prime$.
These are used to define the 4-momentum $w_\mu=p_\mu-p_\mu^\prime$.
The electron mass will be considered as negligible compared to the
muon mass, $m_\mu$, and dropped throughout.
This amounts to discarding terms that are suppressed by a further factor
$m_e^2/m_\mu^2$ compared to the corrections considered here.
The application of dispersion relations leads to introduction of an
auxiliary mass that will be denoted, $M$. It is convenient to introduce
three non-negative real variables, $\rho$, $t$ and $z$, given by
\[
\rho=\frac{m_\pi^2}{m_\mu^2}=1.61395...,\ \ \ \ \ \ \ \
t=-\frac{w^2}{m_\mu^2},\ \ \ \ \ \ \ \
z=\frac{M^2}{m_\mu^2}.
\]
where $m_\pi$ is the mass of the neutral pion.

$\Gamma_0$ will denote the tree-level inverse muon lifetime in the
limit of vanishing electron mass
\begin{equation}
\Gamma_0=\frac{G_F^2 m_\mu^5}{192\pi^3}.
\end{equation}

As usual the Dirac matrices are denoted $\gamma_\mu$ and
$\gamma_{\rm L,R}=(1\pm\gamma_5)/2$ are the left- and right-hand helicity
projection operators respectively.

\section{Hadronic Corrections}

The calculation of the hadronic corrections to muon decay
are greatly facilitated by first
performing a Fierz rearrangement of the Fermi contact interaction term in
the Lagrangian. That being done the effective Feynman diagrams that must
be calculated are shown in Fig.1. Diagram (d) represents the insertion the
muon mass counterterm. Since the electron is taken to be massless its
mass counterterm vanishes. The shaded blob in Fig.1 represents the
subtracted photon self-energy that will be denoted
\begin{equation}
\Pi_{\mu\nu}(q^2)=
(q^2\delta_{\mu\nu}-q_\mu q_\nu)
\left[\Pi_{\gamma\gamma}^\prime(q^2)-\Pi_{\gamma\gamma}^\prime(0)\right].
\end{equation}

The contribution to $\Pi_{\mu\nu}(q^2)$ arising from leptons can
be calculated directly in perturbation theory.  For hadrons the vacuum
polarization can be related via dispersion relations to the hadronic
production cross-section
$\sigma_{\rm had}\equiv\sigma(e^+e^-\rightarrow{\rm hadrons})$
taken from experiments.
In the diagrams of Fig.1 the insertion of the vacuum polarization in
the photon propagator amounts to the replacement
\begin{equation}
\frac{\delta_{\mu\nu}}{q^2-i\epsilon}\longrightarrow
\frac{\delta_{\mu\sigma}}{q^2-i\epsilon}
(q^2\delta_{\sigma\tau}-q_\sigma q_\tau)
\left[\Pi_{\gamma\gamma}^\prime(q^2)-\Pi_{\gamma\gamma}^\prime(0)\right]
\frac{\delta_{\tau\nu}}{q^2-i\epsilon}.
\end{equation}

The terms proportional to $q_\mu q_\nu$ cancel amongst the diagrams
(a)--(c).  Upon substituting of the dispersion integral
representation for the photon vacuum polarization one obtains
\begin{equation}
\frac{\delta_{\mu\nu}}{q^2-i\epsilon}\longrightarrow
\frac{\alpha}{3\pi}\int_{4m_\pi^2}^\infty\frac{dM^2}{M^2}R(M^2)
                       \frac{\delta_{\mu\nu}}{q^2+M^2-i\epsilon}
\label{eq:hadronsubs}
\end{equation}
where $R(M^2)\equiv\sigma_{\rm had}/\sigma_{\rm point}$.
Thus the photon is effectively replaced by a massive vector particle
whose mass is subsequently integrated over.

Taken together the diagrams of Fig.1 are finite and may be calculated
by standard means. This typically involves the reduction
of tensor form factors to
scalar integrals using the methods of Passarino and Veltman
\cite{PassarinoVeltman} and extensions \cite{LERGI,LERGII,LERGIII}.
In practice the reduction was performed using a
Mathematica \cite{Wolfram} implementation of the program
LERG-I \cite{LERGIII} that algebraically reduces
tensor form factors to expressions containing only scalar integrals.
Further details are to be found in the Appendix.

Replacing the photon, as described above, by a massive vector particle
leads to an effective interaction vertex for the $\mu$-$e$ current
that takes the form
\begin{equation}
V_\mu(w^2,M^2)=\left(\frac{\alpha}{\pi}\right)\big\{
i\gamma_\mu\gamma_L F_L(w^2,M^2)
     +p_\mu\gamma_R h_R(w^2,M^2)
     +p_\mu^\prime\gamma_R h_R^\prime(w^2,M^2)\big\}.
\label{eq:EffectiveVertex}
\end{equation}

For a process of the type under scrutiny here the phase space
can be decomposed into a sequence of 2-particle final states
\cite{CzarJezaKuhn}
\[
{\rm dPS}(\mu^-\rightarrow e^-\nu_\mu\bar\nu_e)
\sim dw^2{\rm dPS}(\mu^-\rightarrow e^-w)
         {\rm dPS}(w\rightarrow\nu_\mu\bar\nu_e).
\]
Squaring the matrix element now becomes straightforward especially
since all outgoing fermions are taken to be massless.
One obtains that the change in the inverse
lifetime of the muon induced by an effective interaction
vertex of the form given in Eq.~(\ref{eq:EffectiveVertex}) is
\begin{equation}
\Delta\Gamma(M^2)=\Gamma_0
             \left(\frac{\alpha}{\pi}\right)
     \int_0^1 dt\ 4(1-t)^2\left\{(2t+1)F_L
                  +m_\mu(1-t)\frac{h_R+h_R^\prime}{2}\right\}
\label{eq:widthshift}
\end{equation}
in which
\begin{align}
\int_0^1 dt\
 4(1-t)^2&\left\{(2t+1)F_L
             +m_\mu(1-t)\frac{h_R+h_R^\prime}{2}
       \right\}\notag\\
=&-\frac{1}{432}\left(513+2540z-180z^2-144z^3\right)\notag\\
 &-\frac{1}{72}\left(120+172z-45z^2-12z^3\right)\ln z
 \label{eq:phaseint}\\
 &+\frac{1}{24}\left(132-62z-11z^2+4z^3\right)
        \sqrt{\frac{z}{z-4}}\ln\frac{r_1}{r_2}\notag\\
 &-\frac{1}{6}\left(6+16z-18z^2+z^4\right)
          \left(\frac{\pi^2}{6}+\ln r_1\ln r_2\right)\notag\\
 &+\frac{z}{6}\left(12-2z-z^2\right)
        \sqrt{z(z-4)} \left({\rm Li}_2\frac{1}{r_1}
                           -{\rm Li}_2\frac{1}{r_2}\right)\notag
\end{align}
and the quantities $r_1$ and $r_2$ are defined in the Appendix.
The factor $\sqrt{z-4}$ in the denominator of the third term comes from
the anomalous threshold of a 2-point function that arises in the
mass renormalization of the external muon. The accompanying logarithm
also vanishes at $z=4$ and therefore it does not lead to difficulties.

For large $z$ the right hand side of Eq.~(\ref{eq:phaseint}) can be
expanded to give the asymptotic form
\begin{equation}
 \frac{1}{z}\left(\frac{41}{50}-\frac{2}{5}\ln z\right)
+\frac{1}{z^2}\left(\frac{4321}{432}
                   -\frac{10}{3}\zeta(2)
                   -\frac{97}{36}\ln z\right)
                   +{\cal O}\left( \frac{\ln z}{z^3} \right)
\label{largeMexpansion}
\end{equation}
and so will converge when incorporated into dispersion integrals.
Eq.~(\ref{largeMexpansion}) was also obtained directly using a
large mass expansion technique along the lines of Ref.\cite{MassExp}
which provides a stringent check on all stages of the foregoing
calculation.

The hadronic correction to the muon inverse lifetime may now be obtained
by performing the convolution integral indicated in
Eq.~(\ref{eq:hadronsubs})
\begin{equation}
\Delta\Gamma_{\rm had}=
\frac{\alpha}{3\pi}\int_{4\rho}^\infty\frac{dz}{z}R(m_\mu^2 z)
                       \,\Delta\Gamma(z)
\label{eq:hadcorrz}
\end{equation}

The range of integration can be made finite and all radicals
eliminated by the substitution $u=\sqrt{1-4/z}$ which leads to
\begin{equation}
\Delta\Gamma_{\rm had}=
\Gamma_0
\left(\frac{\alpha}{\pi}\right)^2
\int_{\sqrt{1-\rho^{-1}}}^1
R\left(\frac{4m_\mu^2}{1-u^2}\right)\,K(u)\,du
\label{eq:hadcorru}
\end{equation}
where
\begin{align}
K(u)=\frac{u}{9(1-u^2)^4}\bigg\{&
         \frac{1}{72}(1423+18979u^2-11699u^4+513u^6)\notag\\
    -&\frac{2}{3}(85+127u^2-131u^4+15u^6)\ln\frac{1-u^2}{4}\notag\\
    -&\frac{(9-69u^2-37u^4+33u^6)}{u}\ln\frac{1+u}{1-u}
\label{eq:kernel}\\
    -&\frac{2(19+180u^2-30u^4-44u^6+3u^8)}{(1-u^2)}\notag\\
  &\ \ \ \ \ \ \ \ \ \ \ \ \ \ \ \ \times\left(\frac{\pi^2}{6}
         +\ln\frac{1-u}{2}\ln\frac{1+u}{2}\right)\notag\\
    +&\frac{64u(3+4u^2-3u^4)}{(1-u^2)}
         \left({\rm Li}_2\frac{1+u}{2}-{\rm Li}_2\frac{1-u}{2}\right)
                        \bigg\}.\notag
\end{align}

The expression for $K(u)$ in Eq.(\ref{eq:kernel}) obviously suffers
from strong numerical cancellations as $u\rightarrow1$. In this region
$K(u)$ is best calculated by series expansion
\begin{multline}
K(u)=\left\{\frac{41}{300}
           +\frac{1}{18}\left(\frac{85993}{7200}-5\zeta(2)\right)(1-u)
           +{\cal O}\left((1-u)^2\right)\right\}\\
    +\frac{1}{15}\ln\frac{1-u}{2}\left\{1+\frac{341}{144}(1-u)
           +{\cal O}\left((1-u)^2\right)\right\}.
\end{multline}
The convergence of the integral over hadronic data can be improved by
writing Eq.~(\ref{eq:hadcorru}) as
\begin{multline}
\Delta\Gamma_{\rm had}=
\Gamma_0
\left(\frac{\alpha}{\pi}\right)^2
R(\infty)\int_{\sqrt{1-\rho^{-1}}}^1 K(u)\,du\\
+\Gamma_0
\left(\frac{\alpha}{\pi}\right)^2
\int_{\sqrt{1-\rho^{-1}}}^1
\left\{R\left(\frac{4m_\mu^2}{1-u^2}\right)-R(\infty)\right\}\,K(u)\,du
\label{eq:converge}
\end{multline}
The first integral on the right hand side of Eq.~(\ref{eq:converge}) can
be solved exactly but the result involves trilogarithms, ${\rm Li}_3$,
with arguments containing radicals.
However, its numerical value is well-determined and sufficient for
practical purposes
\begin{equation}
\int_{\sqrt{1-\rho^{-1}}}^1 K(u)\,du=-0.0316710.
\end{equation}
The other integral in Eq.~(\ref{eq:converge}) can now be evaluated
numerically by using a suitable parameterization of hadronic data.
The effect of a narrow resonance of mass $M_R$ and a decay width to
$e^+e^-$ of $\Gamma_{e^+e^-}$ is taken into account by representing it
as a suitably normalized Dirac delta function. A resonance of this type
yields a contribution
\begin{equation}
\Delta\Gamma_R=\Gamma_0\frac{18}{\pi}
               \frac{m_\mu^2\Gamma_{e^+e^-}}
                    {M_R^3\sqrt{1-\frac{4m_\mu^2}{M_R^2}}}
              K\left(\sqrt{1-\frac{4m_\mu^2}{M_R^2}}\right).
\label{eq:narrowRes}
\end{equation}
The total hadronic contribution to inverse lifetime of the muon
\footnote{
Using Eq.s (\ref{eq:converge}) and (\ref{eq:narrowRes}) in conjunction
with his own parameterization of the hadronic data Swartz \cite{Swartz}
obtains
$\Delta\Gamma_{{\rm had}}=-\Gamma_0(\alpha/\pi)^2 (0.0413\pm0.0017)$
in good agreement with our result.}
then is
\begin{equation}
\Delta\Gamma_{{\rm had}}=-\Gamma_0\left(\frac{\alpha}{\pi}\right)^2
                            0.042
\end{equation}

A direct and immediate spin off of Eq.~(\ref{eq:hadcorrz})
is that the contributions from the diagrams Fig.1
in which the hadronic vacuum polarization is replaced by a muon or tau
loop can be easily obtained. Electron loops will not be considered here
as they need to be taken together with real $e^+e^-$ production in order
to yield a physically meaningful result.

In the case of muons one writes
\begin{equation}
R(m_\mu^2z)=\left(1+\frac{2}{z}\right)\sqrt{1-\frac{4}{z}}
\end{equation}
and setting $\rho=1$ the integral in Eq.~(\ref{eq:hadcorru}) can be performed
exactly giving
\begin{eqnarray}
\Delta\Gamma_{{\rm muon}}&=&\Gamma_0\left(\frac{\alpha}{\pi}\right)^2
\left(\frac{16987}{576}-\frac{85}{36}\zeta(2)
                        -\frac{64}{3}\zeta(3)\right)\\
                         &=&-\Gamma_0\left(\frac{\alpha}{\pi}\right)^2
                            0.0364333.
\end{eqnarray}

For tau leptons, the decoupling theorem predicts that their contribution will
be suppressed by a factor of $m_\mu^2/m_\tau^2$. Numerical evaluation
of the integral (\ref{eq:hadcorru}) for the case of tau loops
and using $m_\tau^2/m_\mu^2=282.9$ yields
\begin{equation}
\Delta\Gamma_{{\rm tau}}=-\Gamma_0\left(\frac{\alpha}{\pi}\right)^2
0.00058.
\end{equation}
which, as expected, is significantly smaller and can be discarded for
most practical purposes.

\section{Summary and Conclusions}

The hadronic contributions to the muon lifetime were calculated
in the Fermi model using dispersion relations.
These form an independent subclass of the 2-loop QED corrections that
cannot be evaluated using perturbative methods.
In most processes for which the hadronic contributions are known,
the momenta of the external fermions are fixed but for muon decay
the calculation is complicated by the fact that the outgoing electron
participates in the phase space integration.

As a bonus, an exact expression for the contribution due to
muon loops in the photon vacuum polarization and a numerical value for
tau loops were also obtained. Both of these could be calculated by
other means. The present work does not yield the contribution coming
of electron loops as these must be taken together with $e^+e^-$ pair
creation in order to produce an infrared finite result.

The size of the hadronic contribution is found to be very small compared
to the current experimental error and about 1/8 of that anticipated
in the next generation of measurements of the muon lifetime.
The hadronic uncertainty, coming from the inclusion of hadronic data,
is now safely under control.
The hadronic and muon loop contributions taken together constitute
roughly a quarter of a standard deviation shift in the value that
would be extracted for $G_F$. This is small but nonnegligible.
The contribution from tau loops, on the other hand, can be safely
ignored.

The full set of 2-loop QED corrections, of which the hadronic
corrections form a part, when available
will immediately halve the overall error on the value of $G_F$ since
they will drive the theoretical uncertainty, currently the dominant error,
down to a level of a few parts in $10^8$. This is true regardless of
overall size of the corrections.

\section{Appendix}

Following the notation of Passarino and Veltman \cite{PassarinoVeltman},
the general 2-point scalar integral is defined in dimensional
regularization to be
\begin{equation}
B_0(p^2;m_1^2,m_2^2)=
\int\frac{d^nq}{i\pi^2}\frac{1}{[q^2+m_1^2][(q+p)^2+m_2^2]}
\end{equation}
and the general 3-point scalar integral is
\begin{multline}
C_0(p_1^2,p_2^2,p_5^2;m_1^2,m_2^2,m_3^2)=\\
\int\frac{d^nq}{i\pi^2}
    \frac{1}{[q^2+m_1^2][(q+p_1)^2+m_2^2][(q+p_1+p_2)^2+m_3^2]},
\end{multline}
where $p_5=p_1+p_2$.

Evaluating the scalar integrals by of Feynman parameter methods
for the parameters that appear here manifests the quadratic
\[
x^2-zx+z-i\epsilon=0,
\]
for which the roots are
\[
r_1=\frac{z}{2}+\frac{1}{2}\sqrt{z(z-4)}+i\epsilon,\ \ \ \ \ \ \
r_2=\frac{z}{2}-\frac{1}{2}\sqrt{z(z-4)}-i\epsilon.\ \ \ \ \ \ \
\]
The 2-point scalar integrals that occur here are
\begin{eqnarray}
B_0(0;m_\mu^2,m_\mu^2)&=&\Delta-\ln m_\mu^2\\
B_0(0;M^2,M^2)&=&B_0(0;m_\mu^2,m_\mu^2)-\ln z\\
B_0(w^2;0,m_\mu^2)&=&B_0(0;m_\mu^2,m_\mu^2)
                         +\frac{(1-t)}{t}\ln(1-t)+2\\
B_0(-m_\mu^2;m_\mu^2,M^2)&=&B_0(0;m_\mu^2,m_\mu^2)
                          -\frac{z}{2}\ln z
                          +\frac{1}{2}\sqrt{z(z-4)}\ln\frac{r_1}{r_2}
                          +2\ \ \ \ \ {}
\end{eqnarray}
in which $\Delta$ is a logarithmically divergent constant that arises from
the use of dimensional regularization.

The 3-point form factor may be written
\begin{multline}
C_0(-m_\mu^2,w^2,0;M^2,m_\mu^2,0)=\int_0^1\frac{dy}{m_\mu^2y[y-(1-t)]}\\
  \times\left\{\ln[y^2-zy+z-i\epsilon]
              -\ln\left[(1-t-z)y+z-i\epsilon\right]\right\}.
\label{eq:C0intrep}
\end{multline}
As shown Ref.\cite{GKKS}, the integral representation takes this
particularly simple form because one of the internal masses, that of the
electron, is zero. Integrating the right hand side of
Eq.~(\ref{eq:C0intrep}) by methods given in Ref.\cite{tHooftVeltman79}
leads to a result involving several dilogarithms, ${\rm Li}_2$.

The integration with respect to $t$ that is required in the phase
space integration of Eq.~(\ref{eq:widthshift}) is somewhat arduous but
was facilitated by first obtaining the relation
\begin{align}
m_\mu^2\int_0^1 dt\,(1-t)^n
         C_0(-m_\mu^2,&w^2,0;M^2,m_\mu^2,0)=\notag\\
-\frac{1}{n}\Bigg\{
    &-\int_0^1 dt \frac{t^n-1}{t-1} \ln t
     +\frac{1}{n}\ln z\notag\\
    &+\frac{1}{2}\ln z \int_0^1 dt
          \left(r_1\frac{t^{n-1}-r_1^{n-1}}{t-r_1}
               +r_2\frac{t^{n-1}-r_2^{n-1}}{t-r_2}\right)\notag\\
    &+\int_0^1 dt \left( \frac{t^n-r_1^n}{t-r_1}
                       + \frac{t^n-r_2^n}{t-r_2}\right)
                       \ln\frac{t}{z}\\
    &+\left(\frac{r_1^n+r_2^n}{2}-1\right)
          \left(\frac{\pi^2}{6}+\ln r_1\ln r_2\right)\notag\\
    &-\frac{1}{2}\ln\frac{r_1}{r_2}\int_0^1 dt
          \left(r_1\frac{t^{n-1}-r_1^{n-1}}{t-r_1}
               -r_2\frac{t^{n-1}-r_2^{n-1}}{t-r_2}\right)\notag\\
    &+\frac{r_1^n-r_2^n}{2}
          \left({\rm Li_2}\frac{1}{r_1}-{\rm Li_2}\frac{1}{r_2}\right)
      \Bigg\}\notag
\end{align}
valid for integer $n>0$. The integrals that remain involve only simple
polynomials and logarithms with positive arguments.

\section{Acknowledgements}

The authors wish to thank H.\ Burkhardt and B.\ A.\ Kniehl for their
parameterization and computer code for hadronic data and M.\ L.\ Swartz
for providing us with his evaluation of the hadronic contribution
and uncertainty. Helpful discussions with R.~Akhoury and Y.-P.\ Yao
are also gratefully acknowledged. This work was supported in part by
the US Department of Energy.

\end{document}